\documentclass[pra,amsmath,amssymb,amsfonts,nofootinbib]{revtex4}
\usepackage{bm,graphicx,mathrsfs}

\usepackage{amsmath,bbm}
\usepackage{amsfonts,amssymb}

\newcommand{\M}{{\cal M}}
\newcommand{\T}{{\cal T}}
\newcommand{\1}{\mathbbm{1}}
\newcommand{\id}{{\rm id}}
\newcommand{\qed}{}
\newcommand{\tr}{{\rm tr}}

\newcommand{\be}{\begin{equation}}
\newcommand{\ee}{\end{equation}}
\newcommand{\bea}{\begin{eqnarray}}
\newcommand{\eea}{\end{eqnarray}}

\newcommand{\avr}[1]{\langle#1\rangle}

\newcommand{\ra}{\rightarrow}
\newcommand{\HS}{{\mathfrak{H}}}

\newcommand{\x}{\rangle\langle}
\newcommand{\p}{{\mathfrak{P}}}
\newcommand{\tf}{{\mathfrak{T}}}
\begin{document}

\newtheorem{theorem}{Theorem}
\newtheorem{lemma}[theorem]{Lemma}
\newtheorem{corollary}[theorem]{Corollary}
\newtheorem{proposition}[theorem]{Proposition}
\newtheorem{definition}[theorem]{Definition}
\newtheorem{example}[theorem]{Example}

\newenvironment{proof}{\vspace{1.5ex}\par\noindent\textbf{Proof }}%
    {\hspace*{\fill}$\Box$\vspace{1.5ex}\par}
\newenvironment{remark}{\vspace{1.5ex}\par\noindent{\it Remark}}%
    {\hspace*{\fill}$\Box$\vspace{1.5ex}\par}

\title{{ \sc\Large Dividing Quantum Channels}}

\author{Michael M. Wolf, J. Ignacio Cirac}
\affiliation{Max-Planck-Institute for Quantum Optics,
 Hans-Kopfermann-Str.\ 1,\\ 85748 Garching, Germany.}

\date{\today}

\begin{abstract}We investigate the possibility of dividing
 quantum channels into concatenations of other
channels, thereby studying the semigroup structure of the set of
completely-positive trace-preserving maps. We show the existence
of `\emph{indivisible}' channels which can not be written as
non-trivial products of other channels and study the set of
`\emph{infinitesimal divisible}' channels which are elements of
continuous completely positive evolutions. For qubit channels we
obtain a complete characterization of the sets of indivisible and
infinitesimal divisible channels. Moreover, we identify those
channels which are solutions of time-dependent master equations
for both positive and completely positive evolutions. For
arbitrary finite dimension we prove a representation theorem for
elements of continuous completely positive evolutions based on new
results on determinants of quantum channels and Markovian
approximations.
\end{abstract}

\maketitle  \tableofcontents

\newpage\section{Introduction}

Completely positive linear maps describe the dynamics of a quantum
system in all cases where the evolution is independent of the past
of the system. In the realm of quantum information theory these
maps are referred to as \emph{quantum channels} \cite{Holevo} and,
clearly, the concatenation of two quantum channels is again a
 quantum channel. In this paper we address the converse and
investigate whether and how a channel can be expressed as a
non-trivial concatenation of other channels. That is, we study the
semigroup structure of the set of quantum channels whose input and
output systems have equal finite dimension. Despite the fact that
one-parameter semigroups of completely positive maps are
extensively studied since the late sixties, the semigroup
structure of the set of quantum channels as a whole appears to be
widely unexplored.

\begin{figure}[t]
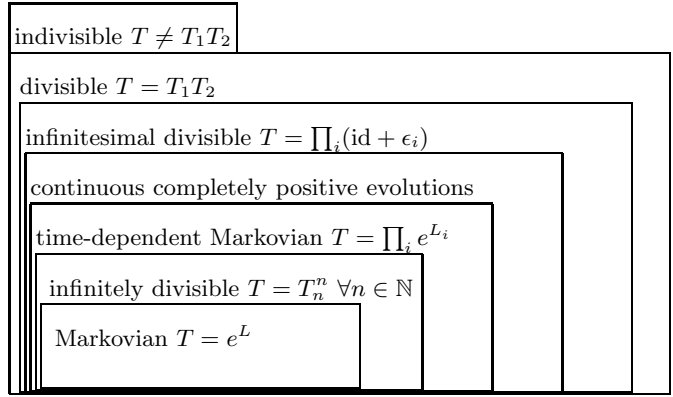

  \begin{center}\begin{tabular}{|c|c|}\cline{1-1}\rule{0pt}{4ex}indivisible $T\neq T_1T_2$\\ \hline \multicolumn{2}{|c|}{\begin{tabular}{p{8.5cm}} \rule{0pt}{4ex}divisible $T=T_1T_2$\\
\begin{tabular}{|p{8cm}|}\hline
\rule{0pt}{4ex}infinitesimal divisible $T=\prod_i(\id+\epsilon_i)$\\
\begin{tabular}{|p{7cm}|}\hline
\rule{0pt}{4ex}continuous completely positive evolutions\\
\begin{tabular}{|p{6cm}|}\hline
\rule{0pt}{4ex}time-dependent Markovian $T=\prod_i e^{L_i}$\\
\begin{tabular}{|p{5cm}|}\hline
\rule{0pt}{4ex} infinitely divisible $T= T_n^n\ \forall n\in\mathbb{N}$\\
\begin{tabular}{|p{4.1cm}|}\hline
\rule{0pt}{4ex} Markovian $T= e^{L}$\\
  \\
  \hline\end{tabular}
  \\
  \hline\end{tabular}
  \\
  \hline\end{tabular}
  \\
  \hline\end{tabular}
  \\
  \hline\end{tabular}
 \end{tabular}} \\
  \hline\end{tabular}
     \end{center}
     \caption{Graphical depiction of the set of quantum channels regarding finer and coarser notions of divisibility, i.e, the possibility
     of expressing a channel in terms of a concatenation of other channels. Whereas the set of \emph{Markovian} channels only contains
     elements of completely positive semigroups, the set of \emph{divisible} channels only requires the existence of any non-trivial product decomposition.
     Definitions of the sets are given in the text (Secs.\ref{sec:Div},\ref{sec:invdiv}). Indivisible maps
     are discussed in Secs.\ref{sec:Div},\ref{sec:qubit} and in Sec.\ref{sec:invdiv}
     it is shown that the sets of infinitesimal divisible and time-dependent Markovian channels coincide.}
\end{figure}

The main purpose of our work is to classify the set of quantum
channels with respect to (i) whether a division in terms of a
concatenation is at all possible and (ii) whether a channel allows
for a division into a large number of infinitesimal channels. This
will lead us to the notions of \emph{divisibility} and
\emph{infinitesimal divisibility}, where the latter property is
equivalent to the existence of a continuous time-dependent
 completely positive evolution which has the given channel as endpoint.
 This classification will allow us to identify basic building blocks (generators) from which all channels can be obtained by
concatenation. Furthermore, it helps us to identify those channels
which are solutions of time-dependent master equations.

A graphical depiction of different notions of divisibility and
their relations is given in Fig.1. The following gives an overview
on the paper and a simplified summary of the obtained results:
\begin{itemize}
    \item Sec.\ref{sec:pre} introduces basic results and provides
    a coarse but later on very useful Markovian approximation to any quantum
    channel.
    \item In Sec.\ref{sec:det} we prove some properties of the
    determinant of quantum channels. In particular, its strict
    monotonicity under concatenation, continuity bounds and properties for Kraus
    rank-two channels and Markovian channels.
    \item The notions of divisible and indivisible maps are
    introduced in Sec.\ref{sec:Div}. The existence of indivisible
    maps and generic divisibility is shown in any dimension, and it is
    proven that building equivalence classes under filtering
    operations preserves divisibility.
    \item Sec.\ref{sec:invdiv} shows that every infinitesimal
    divisible
    channel can be written as a product of Markovian channels and
    that infinitesimal divisibility is preserved under invertible
    filtering operations. Equivalence to the set of continuous
    completely positive evolutions is proven.
    \item Sec.\ref{sec:qubit} provides a complete characterization
    of divisible and indivisible qubit channels in terms of their
    Lorentz normal form. Solutions of time-dependent master equations are identified for both positive and completely positive evolutions.
     Channels with Kraus rank two are studied
    in greater detail separately.
    \item It is shown that already in the qubit case the vicinity
    of the ideal channel contains all types of channels, in
    particular ones that are not infinitesimal divisible and even
    indivisible ones.
\end{itemize}

Before going into detail we want to briefly mention some related
fields and results. The notion \emph{infinite divisibility} goes
back to de Finetti and has thus its origin in classical
probability theory where it means that for any $n\in\mathbb{N}$ a
characteristic function $\chi$ is a power of another
characteristic function $\chi=\chi_n^n$. Examples are the normal
and Poisson distribution. Similarly, the notion of
\emph{indecomposable distributions} exists for those that cannot
be represented as the distribution of the sum of two non-constant
independent random variables.

In the `non-commutative' context, \emph{infinite divisibility} of
positive matrices with respect to the Hadamard product was studied
by Horn \cite{Horn} and the notion was extended to quantum
measurements and quantum channels by Holevo \cite{Hinf} and
Denisov \cite{Denisov}. In fact, the findings of Horn can also be
translated to the quantum world when considering channels with
diagonal Kraus operators, as those act on a density operator by a
Hadamard product with a positive matrix.

\section{Preliminaries}\label{sec:pre}
This section introduces the notation and recalls some basic
results which we will need in the following. Throughout we will
consider linear maps $T:\M_d\ra\M_d$ from the space $\M_d$ of
$d\times d$ matrices into itself. It will be convenient to
consider $\M_d$ as a Hilbert space $\HS_d$ equipped with the
Hilbert-Schmidt scalar product $\avr{A,B}_{\HS}=\tr[A^\dagger B]$
and the 2-norm $||A||_2=\sqrt{\avr{A,A}_\HS}$.\footnote{In general
the Schatten $p$-norms are denoted by $||A||_p:=\big(
\tr[|A|^p]\big)^{1/p}$.} Since $\HS_d$ is isomorphic to
$\mathbb{C}^{d^2}$ the space of linear maps on $\M_d$
(\emph{Liouville space}) is in turn isomorphic to $\M_{d^2}$.
Eigenvalues and singular values of the map $T$ are then understood
as the respective quantities of the matrix representation
$\hat{T}\in \M_{d^2}$ of $T$. More explicit
$\hat{T}_{\alpha,\beta}:= \tr\big[F_\alpha^\dagger
T(F_\beta)\big]=\avr{F_\alpha|T|F_\beta}_\HS$, where
$\{F_\alpha\}_{\alpha=1\ldots d^2}$ is any orthonormal basis in
$\HS_d$. Depending on convenience we will use three different
bases: (i) matrix units $\{|i\x j|\}_{i,j=1\ldots d}$, (ii)
generalized Gell-Mann matrices which are either diagonal or an
embedding of the Pauli matrices $\sigma_x/\sqrt{2}$ and
$\sigma_y/\sqrt{2}$ in $\M_d$, or (iii) a normalized unitary
operator basis given by \be\label{Funi}F_\alpha =
\frac{U_{\alpha_1,\alpha_2}}{\sqrt{d}},\quad
U_{\alpha_1,\alpha_2}=\sum_{r=0}^{d-1} e^{2\pi i r \alpha_2/d}
|\alpha_1+r\x r|\;,\quad \alpha_1,\alpha_2=0,\ldots,d-1\;.\ee When
considering $T$ as a linear map on $\HS_d$ the natural norm is
given by \be
||T||:=\sup_{||A||_2=1}||T(A)||_2=||\hat{T}||_\infty\;.\ee

We denote by $\p$ and $\p^+$ the sets of linear maps on $\M_d$
which are positive and completely positive respectively. The
corresponding subsets of trace preserving maps will be denoted by
$\tf$, $\tf^+$ and the elements of the latter are called
\emph{channels} (in the Schr\"odinger picture). Following
Jamiolkowski's state-channel duality \cite{Jami,Choi} we can
assign to every channel $T$ a state (density operator) $\tau$ by
acting with $T$ on half of a maximally entangled state
$\omega=\frac1{d}\sum_{i,j=1}^d|ii\x jj|$:\be \tau=
(T\otimes\id)(\omega).\ee The rank of this \emph{Jamiolkowski
state} $\tau$ (the un-normalized form of which is often called
\emph{Choi matrix}) is equal to the \emph{Kraus rank} of $T$,
i.e., the minimal number of terms in a Kraus representation
\cite{Kraus} $T(A)=\sum_\alpha K_\alpha A K_\alpha^\dagger$.
Moreover, with the involution
$\avr{ij|\tau^\Gamma|kl}:=\avr{ik|\tau|jl}$ the matrix $
\tau^\Gamma$ leads to a matrix representation of $T$ (with matrix
units as chosen basis \cite{WPG}) such that \be\label{eq:tauGamma}
\hat{T}=d \tau^\Gamma=\sum_\alpha K_\alpha\otimes
\bar{K}_\alpha\;.\ee By $T^*$ we will denote the \emph{dual} of a
map $T$ defined by
 $\tr[T^*(A)B]=\tr[A T(B)]$. If $T$ is trace-preserving then $T^*$
 is \emph{unital}, i.e., $T^*(\1)=\1$ and  the matrix
 representation corresponding to $T^*$ is given by the adjoint
 $\hat{T}^\dagger$.

 A channel will be called \emph{Markovian} if it is an element of a
 completely positive continuous one-parameter semigroup. That is, there exists a generator $L:\M_d\ra\M_d$ with $L^*(\1)=0$ such that  $T_t=e^{t
 L} \in \tf^+$ for all $t\geq 0$. Two equivalent standard forms for
 such generators were derived in \cite{Lindblad} and \cite{Gorini}:
 \bea
 L(\rho) &=& i [\rho,H] + \sum_{\alpha,\beta}
 G_{\alpha,\beta} \left(F_\alpha\rho F_\beta^\dagger -\frac12\{F_\beta^\dagger
 F_\alpha,\rho\}_+\right)\label{normal1}\\
 &=& \label{normal2}i [\rho,H]+ \phi(\rho) - \frac12\{\phi^*(\1),\rho\}_+\;,
 \eea
 where  $G\geq
 0$, $H=H^\dagger$ and $\phi\in\p^+$. The decomposition of the generator $L$
 into a \emph{Hamiltonian part} ($i [\cdot,H]$) and a \emph{dissipative part} ($L-i [\cdot,H$) becomes unique \cite{Gorini} if
 the sum in Eq.(\ref{normal1}) runs only over traceless operators
 $(\tr [F_\gamma]=0)$ from an orthonormal basis in $\HS_d$ (e.g., the one in Eq.(\ref{Funi})). We will in the following always understand Eq.(\ref{normal1}) in this form and call $L$ and the corresponding semigroup \emph{purely dissipative}
 if   $H=0$ w.r.t. such a representation.

 Clearly, not every channel is Markovian (cf. \cite{snapshot}). However, the following  Lemma allows us to
 assign a semigroup to each channel:
 \begin{lemma}[Markovian approximation]\label{lemma:Mapprox}
 For every channel $T\in\tf^+$ we have that
 $e^{t(T-\id)}$, $t\geq 0$ is a completely positive semigroup. Moreover,  if $U_0$ is the unitary conjugation\footnote{By \emph{unitary conjugation} we mean a channel of the form $\rho\mapsto V\rho V^\dagger$ with $V$ being a unitary.} for which the supremum
 $\sup_U\tr_\HS[TU]$ is attained, then $(TU_0-\id)$ is the generator of a purely dissipative semigroup.
 \end{lemma}
\begin{proof}
We will first show that $(T-\id)$ is a valid generator by bringing
into the form of Eq.(\ref{normal2}). Define $\phi(\rho):=
\sum_\alpha A_\alpha \rho A_\alpha^\dagger$ with Kraus operators
$A_\alpha=K_\alpha-x_\alpha\1$ where $\{K_\alpha\}$ are the Kraus
operators of $T$ and $x$ is any unit vector. Then
\be\label{generatorKappa} (T-\id)(\rho) = \phi(\rho) +\kappa\rho
+\rho\kappa^\dagger,\ee with $\kappa=\sum_\alpha \bar{x}_\alpha
A_\alpha$. The trace preserving property $T^*(\1)=\1$ imposes that
$\phi^*(\1)+\kappa+\kappa^\dagger=0$ so that the Hermitian part of
$\kappa$ is $-\phi^*(\1)/2$. If we denote by $-i H$ with
$H=H^\dagger$ the anti-Hermitian part then $\kappa=-\phi^*(\1)/2-i
H$ which leads to the form in Eq.(\ref{normal2}), proving the
first statement. Note that there is  freedom in the choice of the
anti-Hermitian part of $\kappa$ as Eq.(\ref{generatorKappa}) is
invariant under adding to $\kappa$ a multiple of $i\1$.

In order to prove the second statement we have to exploit the
freedom \cite{Davies} in the decomposition into dissipative and
Hamiltonian part, where the latter corresponds (up to multiples of
$i\1$) to the anti-Hermitian part of $\kappa$. Note that \bea
(T-\id)(\rho) &=& \sum_\alpha\label{eq:AKraus}
(A_\alpha-a_\alpha\1)\rho(A_\alpha-a_\alpha\1)^\dagger +
\kappa_a\rho +\rho\kappa_a^\dagger\;,\\
\kappa_a&=& \kappa+\sum_\alpha \bar{a}_\alpha A_\alpha
-\1|a_\alpha|^2/2\\
&=& \sum_\alpha
(\bar{x}_\alpha+\bar{a}_\alpha)(K_\alpha-x_\alpha\1)-\1|a_\alpha|^2/2\label{secondkappa}
\eea gives other representations of the same generator for any
complex vector $a$. The representation of the generator has
traceless Kraus operators in Eq.(\ref{eq:AKraus}) iff
$({x}_\alpha+{a}_\alpha)=\tr[K_\alpha]/d$. For this choice of $a$
the imaginary part of $\kappa_a$ in Eq.(\ref{secondkappa}) would
thus indeed be a multiple of $i\1$ if $\sum_\alpha
\tr[\bar{K}_\alpha]K_\alpha\geq 0$. Let us now show that exactly
this is achieved by concatenating $T$ with $U_0$. To this end note
 that by exploiting Eq.(\ref{eq:tauGamma}) we get \bea \label{eq:2V}\tr_\HS[TU_0]&\leq& \sup_{V,V'} \Big|\sum_\alpha
\tr\overline{[K_\alpha V']}\tr[K_\alpha V]\Big|\\ &=&
\sup_{V,V'}\big|\avr{\phi_{V'}|\tau|\phi_V}\big|\leq\sup_V||\sqrt{\tau}|\phi_V\rangle||^2\\
&=& \tr_\HS[TU_0],\label{eq:3V}\eea where $V,V'$ are unitaries and
$|\phi_V\rangle=\sqrt{d}(V\otimes\1)\sum_{i=1}^d|ii\rangle$. On
the one hand the r.h.s. of Eq.(\ref{eq:2V}) is maximized if $V$
and $V'$ are unitaries from the polar decomposition of the
remaining parts, i.e., $V$ for instance is the polar unitary of
$\sum_\alpha \tr\overline{[K_\alpha V']} K_\alpha$. On the other
hand it follows from equality to $\tr_\HS[TU_0]$ that the maximum
is attained for $V=V'$ so that for $U_0(\cdot)=V\cdot V^\dagger$
we get $\sum_\alpha \tr\overline{[K_\alpha V]}K_\alpha V\geq 0$
concluding the proof.
\qed
\end{proof}

\section{Determinants}\label{sec:det}

The multiplicativity property of determinants $\det(T_1 T_2)=(\det
T_1)(\det T_2)$ makes them an indispensable  tool for the study of
semigroup properties of sets of linear maps. The following theorem
contains some of their basic properties. Though the results of
this section are necessary for subsequent proofs they are not
essential for understanding the parts on divisibility, so that
this section might be skipped by the reader.

\begin{theorem}[Determinants]\label{ThmDet}
Let $T:\M_d\ra\M_d$ be a linear positive and trace preserving map.
\begin{enumerate}
    \item $\det T$ is real and
contained in the interval $[-1,1]$,
    \item $|\det T|=1$ iff $\;T$
is either a unitary conjugation or unitarily equivalent to a
matrix transposition,
    \item if $T$ is a unitary conjugation then $\det T=1$ and if $\det T=-1$ then $T$ is a matrix transposition up to unitary equivalence.
    In both cases the
    converse holds iff $\lfloor\frac{d}2\rfloor$  is odd.
\end{enumerate}
\end{theorem}
\begin{proof}
First note that every positive linear map satisfies
$T(A^\dagger)=T(A)^\dagger$ for all $A\in\M_d$. This becomes
obvious by writing $A$ as a linear combination of four positive
matrices and using linearity of $T$. As a consequence all
eigenvalues either come in complex conjugate pairs or are real so
that $\det T$ is real. From the boundedness of the norm of any
trace preserving $T\in\tf$ ($||T||\leq \sqrt{d}$ \cite{PWPR06})
together with the fact \cite{Bhatia} that the spectral radius
equals $\lim_{m\ra\infty} ||T^m||^{1/m}$ it follows that the
spectral radius is one which implies
 $\det T\in[-1,1]$.

Now consider the case $\det T=\pm1$ where all eigenvalues are
phases. There is always a sequence $n_i$ such that the limit of
powers $\lim_{i\rightarrow\infty} T^{n_i}=:T_\infty$ has
eigenvalues which all converge to one \footnote{This is an
immediate consequence of Dirichlet's theorem on Diophantine
approximations \cite{Dio}.}. To see that this implies that
$T_\infty=\id$ consider a two-by-two block on the diagonal of
the Schur decomposition of $\hat{T}^{n_i}$. Up to a phase this block is of the form $\left(%
\begin{array}{cc}
  1 & c \\
  0 & e^{i\epsilon} \\
\end{array}%
\right)$. Thus by taking the $p$'th power of $T^{n_i}$ this is
mapped to $$
\left(%
\begin{array}{cc}
  1 & \sum_{k=0}^{p-1} e^{ik \epsilon} c \\
  0 & e^{ip \epsilon} \\
\end{array}%
\right). $$ As $\epsilon\rightarrow 0$ for $n_i\rightarrow\infty$
the norm $||(T^{n_i})^p||$ could be increased without limit (by
increasing $p$ with $n_i$) unless $c\ra 0$. However, all powers of
$T$ are trace preserving and positive and have therefore to have
bounded norm. This rules out the survival of Jordan block-like
off-diagonal elements
 so that $T_\infty=\id$. Hence, the inverse
$T^{-1}=T_\infty T^{-1}=\lim_{i\rightarrow\infty} T^{n_i-1}$ is a
trace preserving positive map as well.

Assume that the image of any pure state $\Psi$ under $T$ is mixed,
i.e., $T(\Psi)=\lambda\rho_1+(1-\lambda)\rho_2$ with
$\rho_1\neq\rho_2$. Then by applying $T^{-1}$ to this
decomposition we would get a nontrivial convex decomposition for
$\Psi$ (due to positivity of $T^{-1}$) leading to a contradiction.
Hence, $T$ and its inverse map pure states onto pure states.
Furthermore, they are unital, which can again be seen by
contradiction. So assume $T(\1)\neq\1$. Then the smallest
eigenvalue of $T(\1)$ satisfies $\lambda_{min}<1$ due to the trace
preserving property. If we denote by $|\lambda\rangle$ the
corresponding eigenvector, then
$\1-\frac{\lambda_{min}+1}{2}T^{-1}(|\lambda\rangle\langle\lambda|)$
is a positive operator, but its image under $T$ would no longer be
positive. Therefore we must have $T(\1)=\1$.

Every unital positive trace preserving map is contractive with
respect to the Hilbert-Schmidt norm \cite{PWPR06,Streater}. As
this holds for both $T$ and $T^{-1}$ we have that $\forall
A\in\M_d:||T(A)||_2=||A||_2$, i.e., $T$ acts unitarily on the
Hilbert Schmidt Hilbert space. In particular, it preserve the
Hilbert Schmidt scalar product
$\tr{\big[T(A)T(B)^\dagger\big]}=\tr{[AB^\dagger]}$. Applying this
to pure states $A=|\phi\rangle\langle\phi|$ and
$B=|\psi\rangle\langle\psi|$ shows that $T$ gives raise to a
mapping of the Hilbert space onto itself which preserves the value
of $|\langle\phi|\psi\rangle|$. By Wigner's theorem
\cite{Wigner,Bargmann} this has to be either unitary or
anti-unitary. If $T$ is a unitary conjugation then $\det
T=\det(U\otimes\bar U)=1$. Since every anti-unitary is unitarily
equivalent to complex conjugation, we get that $T$ is in this case
a matrix transposition $T(A)=A^T$ (up to unitary equivalence). The
determinant of the matrix transposition is easily seen in the
Gell-Mann basis of $\M_d$. That is, we take basis elements
$F_\alpha$ of the form ${\sigma_x}/\sqrt{2},{\sigma_y}/\sqrt{2}$
for $\alpha=1,\ldots d^2-d$ and diagonal for
$\alpha=d^2-d+1,\ldots,d^2$. In this basis matrix transposition is
diagonal and has eigenvalues $1$ and $-1$ where the latter appears
with multiplicity $d(d-1)/2$. This means that matrix transposition
has determinant minus one iff $d(d-1)/2$ is odd, which is
equivalent to $\lfloor\frac{d}2\rfloor$ being odd.\qed
\end{proof}

From this we get the following important corollary:
\begin{corollary}[Monotonicity of the determinant]\label{cor:monotonicity}
Consider the set $\tf$ of positive trace preserving linear maps on
$\M_d$. \begin{enumerate}
    \item $T,T^{-1}\in\tf$ iff $T$ is a unitary
conjugation or matrix transposition.
    \item The determinant of $T\in\tf$ is decreasing in magnitude under
    composition, i.e., $|\det T|\geq |\det T T'|$ for all $T'\in\tf$  where
    equality holds iff $T'$ is a unitary, a matrix transposition or $\det T=0$.
\end{enumerate}
\end{corollary}
Part 1. of this corollary is a simple consequence of Wigner's
theorem and was proven for completely positive maps for instance
in \cite{POVMs}.\footnote{It is also a consequence of
\cite{Kadison} and therefore sometimes called \emph{Wigner-Kadison
theorem}.}

 One might wonder whether completely positive maps can
have negative determinants. The following simple example answers
this question in the affirmative. It is build up on the map
$\rho\mapsto \rho^{T_c}$ which transposes the corners of
$\rho\in\M_d$, i.e., $(\rho^{T_c})_{k,l}$ is $\rho_{l,k}$ for the
entries $(k,l)=(1,d),(d,1)$ and remains $\rho_{k,l}$ otherwise.
Note that for $d=2$ this is the ordinary matrix transposition.
\begin{example}\label{exp1}
The map $T:\M_d\rightarrow\M_d$ defined by \be
T(\rho)=\frac{\rho^{T_c}+\1\tr\rho}{1+d}\label{paradigm}\ee is
trace preserving, completely positive with Kraus rank $d^2-1$ and
has determinant $\det T= -\big(d+1\big)^{1-d^2}$. For $d=2$ the
channel is entanglement breaking and can be written as\be
T(\rho)=\frac13\sum_{j=1}^6
|\bar{\xi}_j\rangle\langle\xi_j|\rho|\xi_j\rangle\langle\bar{\xi}_j|,\ee
where the six $\xi_j$ are the normalized eigenvectors of the three
Pauli matrices.
\end{example}
\begin{proof}A convenient matrix representation of the channel is
given in the generalized Gell-Mann basis. Choose $F_1$ as the
$\sigma_y/\sqrt{2}$ element corresponding to the corners and
$F_2=\1/\sqrt{d}$ the only  element which is not traceless. Then
$\hat{T}={\rm diag}[-1,1+d,1,\ldots,1]/(d+1)$ leading to $\det
T=-(d+1)^{1-d^2}$.

 For complete positivity we have to check
positivity of the Jamiolkowski state $\tau$. The corner
transposition applied to a maximally entangled state leads to one
negative eigenvalue $-1/d$. This is, however, exactly compensated
by the second part of the map such that $\tau\geq 0$ with rank
$d^2-1$.

 The representation for $d=2$ is obtained from $\tr[AT(B)]=\tr[(A\otimes B^T)\tau]d$
by noting that in this case $\tau$ is proportional to the
projector onto the symmetric subspace which in turn can be written
as $\frac12\sum_j |\xi_j\rangle\langle\xi_j|^{\otimes 2}$ in
agreement with the given Kraus representation of the channel.\qed
\end{proof}

The above example has Kraus rank $d^2-1$. Channels of Kraus rank
two and compositions thereof can never lead to negative
determinants:
\begin{theorem}[Kraus rank two maps]
All linear maps on $\M_d$ which are completely positive with Kraus
rank two have non-negative determinant. Hence every composition of
such Kraus rank two maps has non-negative determinant.
\end{theorem}
\begin{proof}Let $A$ and $B$ be two Kraus operators of the map $T$
and assume for the moment that $\det A\neq 0$. Then, using matrix
units as a basis for the Hilbert-Schmidt Hilbert space, we can
represent the channel by the matrix
$A\otimes\bar{A}+B\otimes\bar{B}$. If we use the singular value
decomposition of $A=USV$ we can write the determinant as $\det
T=(\det S)^{2d} \det(\1+B'\otimes\bar{B}')$ with
$B'=S^{-1/2}U^\dagger BV^\dagger S^{-1/2}$. Denoting the
eigenvalues of $B'$ by $b_k$ we obtain $\det T=(\det
S)^{2d}\prod_{k,l}(b_k\bar{b}_l+1)=(\det S)^{2d}\prod_{k<
l}|b_k\bar{b}_l+1|^2\prod_j(|b_j|^2+1)$ which is indeed positive.
As the set of maps with $\det A\neq 0$ is dense and $\det T$
continuous we obtain $\det T\geq 0$ for all Kraus rank two
maps.\qed
\end{proof}

For Markovian channels the determinant can easily be expressed in
terms of the dissipative part of the generator:

\begin{theorem}[Determinants of Markovian Channels]\label{thm:detMarkov}
Let $T=e^L\in\tf^+$ be a Markovian channel with generator $L$ of
the form in Eq.(\ref{normal1}). Then $\det T=\exp(-d\;\tr[G])$ and
if $L$ is purely dissipative then $||L||\leq -\frac2d\ln\det T$.
\end{theorem}
\begin{proof}
First note that $\det e^L=e^{\tr L}$ where the trace is understood
in $\HS_d$. A convenient basis for computing this trace is the one
in Eq.(\ref{Funi}) as it is unitary (up to a factor) and the
unitaries fulfill the \emph{Weyl relations}
$U_{a,b}U_{c,d}=U_{a+c,b+d} \exp(2\pi i bc/d)$. The Hamiltonian
part of the generator does not contribute to the trace. For the
dissipative part assume first the basis in Eq.(\ref{normal1}) is
 the one from Eq.(\ref{Funi}) as well. Then straight forward calculation
exploiting unitarity and the Weyl relations gives
$\tr[L]=-d\tr[G]$. Due to the basis independence of the trace this
has to be true independent of the choice of the $\{F_\gamma\}$ in
Eq.(\ref{normal1}) (as long as they are traceless and
orthonormal).

For the second statement in the theorem we use the basis in which
$G={\rm diag}[g_1,g_2,\ldots]$ is diagonal: \be L(\rho)=
\sum_\gamma g_\gamma \Big(F_\gamma\rho F_\gamma^\dagger -\frac12
\{F_\gamma^\dagger F_\gamma,\rho\}_+\Big).\ee From the triangle
inequality together with the fact that $||F_\gamma||_2=1$ we
obtain then $||L||\leq 2\sum_\gamma g_\gamma=2\tr[G]$. \qed

\end{proof}

Thm. \ref{ThmDet} shows that if for a channel $\det T=1$, then it
has to be a unitary. By continuity a channel close to a unitary
will still have determinant close to one. The following is a
quantitative version of the converse: if the determinant is large,
then there is a unitary conjugation (namely the inverse of the one
maximizing $\tr_\HS[TU]$) close to the channel. Similarly, a large
determinant implies a large purity $\tr[\tau^2]$ of the
Jamiolkowski state $\tau$.

\begin{theorem}[Bounds on the determinant]\label{thm:detBounds}
Let $T\in\tf^+$ be a channel on $\M_d$. The purity of the
respective Jamiolkowski state $\tau$ leads to an upper bound on
the determinant \be\label{puritybound}\det T\leq
\tr\big[\tau^2\big]^{d^2/2},\ee and in the limit $\det
T\rightarrow 1$ the distance between $T$ and the  unitary
conjugation $U_0^{-1}$ appearing in Lemma \ref{lemma:Mapprox}
vanishes as
\be\label{scaling}||T-{{U_0}^{-1}}||=O\big(\sqrt{1-\det
T}\big).\ee
\end{theorem}
\begin{remark}
An explicit however lengthy bound for the norm distance in
Eq.(\ref{scaling}) can easily be deduced from the subsequent
proof.
\end{remark}
\begin{proof} To relate the purity to the determinant we exploit
that by Eq.(\ref{eq:tauGamma}) $\hat{T}=d \tau^\Gamma$ is a matrix
representation of the channel, so that
\be\tr[\tau^2]=\frac1{d^2}\tr[\hat{T}^\dagger
\hat{T}]=\frac1{d^2}\sum_{i=1}^{d^2}s_i^2,\ee where the $s_i$ are
the singular values of $\hat{T}$ (and thus $T$). From this
Eq.(\ref{puritybound}) is obtained via the geometric--arithmetic
mean inequality together with the fact that $|\det T|=\prod_i
s_i$.

We now use the purity bound (\ref{puritybound}) to prove the
scaling in Eq.(\ref{scaling}).  The aim of the following is to
relate first the purity to the largest eigenvalue of $\tau$ and
then the latter to the sought distance
$||T-U_0^{-1}||=||TU_0-\id||$.

The largest eigenvalue $\mu:=||\tau||_\infty$ of the Jamiolkowski
state is for $\tr[\tau^2]\geq1/2$ lower bounded by the purity via
\be\mu\geq\frac12\Big(1+\sqrt{2\tr[\tau^2]-1}\Big).\label{eq:12}\ee
If we denote by $\Psi=|\psi\x\psi|$ the projector onto the
eigenstate  corresponding to $\mu$ then
\be||{\1}/d-\tr_A\Psi||_1\leq||\tau-\Psi||_1\leq
2(1-\mu),\label{eq:47}\ee where the first inequality follows from
the monotonicity of the trace norm distance under the partial
trace. We will now use the bounds between the trace norm distance
and the fidelity \cite{fbounds}
$f(\sigma,\rho):=\tr\sqrt{\rho^{1/2}\sigma\rho^{1/2}}$:
\be\label{eq:fid} 1-f(\rho,\sigma)\leq
\frac12||\rho-\sigma||_1\leq\sqrt{1-f(\rho,\sigma)^2}.\ee By
Uhlmann's theorem \cite{Uhlmann} we have that
$f(\tr_A\Psi,\1/d)=\sup_\Omega |\avr{\Omega|\psi}|$ where the
supremum is taken over all maximally entangled states $\Omega$.
Denote by $\tau_0$ the Jamiolkowski state of $TU_0$ and
$|\Omega_0\rangle=\sum_i|ii\rangle/\sqrt{d}$ so that
$\omega=|\Omega_0\x\Omega_0|$. Then by
Eqs.(\ref{eq:2V}-\ref{eq:3V}) we have that
$\avr{\Omega_0|\tau_0|\Omega_0}=\sup_\Omega\avr{\Omega|\tau|\Omega}$
which is in turn lower bounded by $\mu f(\tr_A\Psi,\1/d)^2$.
Together with Eqs.(\ref{eq:47},\ref{eq:fid}) this gives\be
\avr{\Omega_0|\tau_0|\Omega_0}\geq\mu^3\label{eq:mu3}\;.\ee

Finally, we have to relate the distance between the Jamiolkowski
states $\tau_0$ and $\omega$ to that of the respective channels.
This can be done by exploiting $\tr[AT(B)]=\tr[(A\otimes
B^T)\tau]d$ so that in general
$||T_1-T_2||=d\sup_{A,B}|\tr[(\tau_1-\tau_2)A\otimes B]|$ where
the supremum is taken over all operators with $||A||_2=||B||_2=1$
such that an upper bound is given by $d||\tau_1-\tau_2||_1$. In
this way we get \be ||TU_0-\id||\leq d ||\tau_0-\omega||_1\leq
2d\sqrt{1-\avr{\Omega_0|\tau_0|\Omega_0}}\leq
2d\sqrt{1-\mu^3}\label{eq:23}\;.\ee

The scaling in Eq.(\ref{scaling}) is then obtained by combining
Eqs.(\ref{puritybound},\ref{eq:12},\ref{eq:23}) and expanding
around $\det T=1$.\qed
\end{proof}

\newpage\section{Divisible and Indivisible Maps}\label{sec:Div}

In the following we will apply the above results and study
decompositions of channels in terms of concatenations of other
channels, i.e., the possibility of writing $T\in\tf^+$ as
$T=T_1T_2$, $T_i\in\tf^+$. As the notion \emph{decomposable} is
commonly used in the context of convex decompositions and often
refers to a specific convex decomposition of positive maps
\cite{positivemaps}, we will use the notion \emph{divisible}
instead. Clearly, every channel is divisible in a trivial way
$T=(T U^{-1}) U$, where $U$ is any unitary conjugation. In order
to make the \emph{divisibility} of a channel a non-trivial concept
we thus define it up to unitary conjugation:

\begin{definition}[Divisibility]\label{Def:divisibility}
Consider the set $\T\in\{\tf,\tf^+\}$ of linear trace preserving
positive or completely positive maps from $\M_d$ into itself. We
say that $T\in\T$ is \emph{indivisible} if every decomposition of
the form $T=T_1T_2$ with $T_i\in\T$ is such that one of the $T_i$
has to be a unitary conjugation. $T$ is called \emph{divisible} if
it is not indivisible.
\end{definition}
That this concept is not empty, i.e., that indivisible maps indeed
exist is now a simple consequence of Thm.\ref{ThmDet}:

\begin{corollary}[Indivisible positive maps]
Consider the case where $\lfloor\frac{d}2\rfloor$ is odd. Then the
matrix transposition $\theta:\rho\mapsto \rho^T$, $\rho\in\M_d$ is
indivisible within the set $\tf$ of positive trace preserving maps
on $\M_d$.
\end{corollary}
\begin{proof} Assume that the matrix transposition $\theta$ has a decomposition
$\theta=T_1 T_2$. Then by Thm.\ref{ThmDet}
$\det(T_1)=-\det{T_2}=\pm 1$. Hence, one of the two maps has to be
a transposition and the other a unitary.\qed
\end{proof}
Clearly, this simple observation is reminiscent of the fact that a
reflection can not be expressed in terms of rotations (with
positive determinant). The following is a less trivial example
based on the same idea:

\begin{corollary}[Indivisible completely positive maps]\label{Cor:IndCP} Consider the set $\tf^+$
 of completely positive trace preserving maps on $\M_d$. The
 channel $T_0\in\tf^+$ with minimal determinant, i.e., for which
 $\det T_0=\inf_{T\in\tf^+}\det T$, is indivisible. For $d=2$ we
 have that
 $T_0(\rho)=(\rho^T+\1)/3$ is the channel discussed in
 Exp.\ref{exp1}.\label{tff}
\end{corollary}
\begin{proof} By Exp.\ref{exp1} there are always channels with $\det T < 0$.
 As the set $\tf^+$ of channels on $\M_d$ is compact
there is always a map $T_0$ for which $\inf_{T\in\tf^+}\det T$ is
attained. Now consider a decomposition $T_0=T_1T_2$. Then by
Thm.\ref{ThmDet} and Cor.\ref{cor:monotonicity} either $T_1$ or
$T_2$ has to be unitary.

For the case $d=2$ recall that $T$ can be conveniently represented
in terms of the real $4\times 4$ matrix
$\hat{T}_{ij}:=\tr[\sigma_iT(\sigma_j)]/2$ where the $\sigma_i$s
are
identity and Pauli matrices. In general \be\label{T2} \hat{T}=\left(%
\begin{array}{cc}
  1 & 0\\
  v & \Delta \\
\end{array}%
\right),\ee where $v\in\mathbb{R}^3$, $\Delta$ is a $3\times3$
matrix and $\det\Delta=\det T$. To simplify matters we can
diagonalize $\Delta$ by special orthogonal matrices $O_1\Delta
O_2=\mbox{diag}\{{\lambda_1,\lambda_2,\lambda_3}\}$ corresponding
to unitary operations before and after the channel. Obviously,
this does neither change the determinant, nor complete positivity.
For the latter it is necessary that $\vec{\lambda}$ is contained
in a tetrahedron spanned by the four corners of the unit cube with
$\lambda_1\lambda_2\lambda_3=1$ \cite{unitalqubits,unitalqubits2}.
Fortunately, all these points can indeed be reached by unital
channels ($v=0$) for which this criterion becomes also sufficient
for complete positivity. By symmetry we can restrict our attention
to one octant and reduce the problem to maximizing $p_1p_2p_3$
over all probability vectors $\vec{p}$ yielding
$\vec{p}=(\frac13,\frac13,\frac13)=(\lambda_1,\lambda_2,-\lambda_3)$.
Hence, the minimal determinant is $-(\frac13)^{3}$ and the
corresponding channel can easily be constructed from $\hat{T}$ as
$T:\rho\mapsto\frac13(\rho^T+\1)$. \qed\end{proof} The channels
with minimal determinant lie at the border of $\tf^+$, i.e., they
have reduced Kraus rank. In fact, for channels with full Kraus
rank ($d^2$) one can easily see that they are all divisible:

\begin{theorem}[Divisibility of generic
channels]\label{Thm:GenericDivisibility} If a channel $T\in\tf^+$
on $\M_d$ has Kraus rank $d^2$, then $T$ is divisible.
\end{theorem}
\begin{proof}Note that $T$ has full Kraus rank iff the corresponding Jamiolkowski state $\tau=(T\otimes\id)(\omega)$ has
full rank. Let $T_1\in\tf^+$ be any invertible non-unitary
channel. Then $(T_1^{-1}T\otimes\id)(\omega)$ is still positive if
only $T_1$ is sufficiently close to the identity. Therefore
$T_2:=T_1^{-1}T$ is an admissible channel so that $T=T_1T_2$.
Clearly, one can always choose $T_1$ such (e.g. of Kraus rank two,
or $\det T_1\neq \det T$) that neither $T_1$ nor $T_2$ are
unitary.\qed
\end{proof}
We will now see that in searching for a decomposition of a trace
preserving map $T{=}T_1T_2$ we can essentially drop the trace
preserving constraint on $T_1$ and $T_2$. That is, if there exists
a non-trivial decomposition into non-trace preserving maps, then
there will be one in terms of trace preserving maps as well:
\begin{theorem}\label{thm:nontp}
Let ${{\cal P}\in\{\p,\p^+\}}$  be either the set of positive or
completely positive linear maps on $\M_d$, and ${\cal
T}\in\{\tf,\tf^+\}$ the respective subset of trace preserving
maps. Then for every concatenation $\tilde{T}_1\tilde{T}_2=T$,
$\tilde{T}_i\in{{\cal P}},\; T\in{\cal T}$ with
$\det{\tilde{T}_1^*(\1)}\neq 0$ there exist $T_1,T_2\in{\cal T}$
with Kraus rank\footnote{For positive maps we define the Kraus
rank as the rank of the corresponding Jamiolkowski operator (Choi
matrix).} ${\rm rank}[T_i]={\rm rank}[\tilde{T}_i]$ such that
$T_1T_2=T$.
\end{theorem}
\begin{proof}We will explicitly construct $T_1$ and $T_2$ via their
duals. Due to positivity and the absence of a kernel in
$\tilde{T}_1^*(\1)$ we can find a positive definite matrix $P>0$
which is the square root of $\tilde{T}_1^*(\1)=P^2$. Then
$T_1^*(X):=P^{-1}\tilde{T}_1^*(X) P^{-1}$ fulfills $T_1^*(\1)=\1$
and is thus the dual of a map $T_1\in{\cal T}$. Defining
$T_2^*(X):=\tilde{T}_2^*(PXP)$ we obtain $T_2^*T_1^*=T^*$ so that
indeed $T_1T_2=T$. Moreover, $T_2\in {\cal T}$ since
$T_2^*(\1)=T_2^*T_1^*(\1)=T^*(\1)=\1$. Equality for the Kraus
ranks follows immediately from the fact that $T_i$ and
$\tilde{T}_i$ differ merely by concatenation with an invertible
completely positive Kraus rank-one map.\qed\end{proof} For the
classification of (in-)divisible maps this allows us to restrict
to equivalence classes under invertible filtering operations. In
Sec.\ref{sec:qubit} this reduction will enable us to completely
characterize the set of indivisible qubit channels.

\begin{corollary}[Reduction to normal
form]\label{Cor:NormalformReduction} Let ${\cal
T}\in\{\tf,\tf^+\}$ and $T,\tilde{T}\in{\cal T}$ be related via
$T= T_A\tilde{T}T_B$ where $T_A,T_B\in\p^+$ are invertible
completely positive maps with Kraus rank one. Then $T$ is
divisible iff $\tilde T$ is divisible.
\end{corollary}

\section{Infinitesimal Divisible Channels}\label{sec:invdiv}

In this section we will refine the somewhat coarse notion of
divisibility by asking which channels can be broken down into
\emph{infinitesimal} pieces, i.e., into channels arbitrary close
to the identity. This will lead us to a number of a priori
different sets of channels, depending on the additional structure
which we impose on the infinitesimal constituents. The main result
will then be the equivalence of three of these sets, showing that
the imposed structure is not an additional requirement but rather
emerges naturally.

Let us begin with the most structured and best investigated of
these sets: the set of Markovian channels. Evidently, a Markovian
channel, i.e., an element of a continuous completely positive
one-parameter semigroup is divisible. Furthermore it can be
divided into a large number of equal infinitesimal channels and it
is the solution of a time-independent \emph{master equation}
\begin{equation}
\frac{\partial\rho}{\partial t}=L(\rho)\;,
\end{equation}
with $L$ of the form in Eqs.(\ref{normal1},\ref{normal2}).

Following the terminology used in classical probability theory one
 calls a channel $T$ \emph{infinitely divisible}
\cite{Holevo,Denisov} if for all $n\in\mathbb{N}$ there is another
channel $T_n$ such that $T=T_n^n$. It was shown in \cite{Denisov}
that infinitely divisible channels are all of the form $T=T_0e^L$
where $L$ is a Lindblad generator of the form in
Eq.(\ref{normal1}) and $T_0$ is an idempotent channel satisfying
$T_0L=T_0L T_0$. Hence, an infinitely divisible channel becomes an
element of a continuous completely positive one-parameter
semigroup if $T_0=\id$.

Consider now a more general family, which one might refer to as
\emph{continuous completely positive evolutions}. That is, for
some time interval $[0,t]$ there exists a continuous mapping
$[0,t]\times [0,t]\ra\tf^+$ onto a family of
quantum channels $\{T(t_2,t_1)\}$ 
such that \begin{enumerate}
    \item $T(t_3,t_2)T(t_2,t_1)=T(t_3,t_1)$ for all $0\leq t_1\leq t_2\leq t_3\leq
    t$,
    \item $\lim_{\epsilon\ra0}||T(\tau+\epsilon,\tau)-\id||=0$ for
all $\tau\in[0,t)$.
\end{enumerate} In other words there is a continuous path within $\tf^+$ which
connects the identity with each element of this family and along
which we can move (one-way) by concatenation with quantum
channels. Let us denote by ${\cal J}\subset\tf^+$ the set of all
elements of such continuous completely positive evolutions.
Clearly, this set is included in the following:

\begin{definition}[Infinitesimal divisibility]\label{def:infinitesimal}
Define a set ${\cal I}$ of channels $T\in\tf^+$ with the property
 that for all $\epsilon>0$ there exists a finite set of channels
$T_i\in\tf^+$ such that (i) $||T_i-\id||\leq\epsilon$ and (ii)
$\prod_iT_i =T$. We say that a channel is \emph{infinitesimal
divisible} if it belongs to the closure $\overline{\cal I}$.
\end{definition}
\begin{remark}
Note that every infinitely divisible channel is also infinitesimal
divisible. To see this note that for every idempotent channel
$T_0$ we have that $\big[(1-\epsilon)\id+\epsilon T_0\big]^n$ is
 a product of channels which are $\epsilon$-close to the
identity with convergence to $T_0$ for $n\ra\infty$.
\end{remark}
By continuity and multiplicativity of the determinant we obtain a
simple necessary condition for a channel to be infinitesimal
divisible:
\begin{proposition}\label{prop:detpos}
If a channel $T\in\tf^+$ is infinitesimal divisible, then $\det
T\geq 0$.
\end{proposition}
A similar notion of infinitesimal divisibility can be defined by
introducing a set ${\cal I}'$ analogous to the set ${\cal I}$ with
the additional restriction that all the $T_i\in\tf^+$ have to be
Markovian, i.e., of the form $T_i=e^{L_i}$ with $L_i$ a Lindblad
generator. Clearly, $\overline{{\cal I}'}\subseteq\overline{\cal
I}$ and intuitively the converse should also hold as every channel
close to the identity should be `almost Markovian'. However, the
closer the $T_i$ are to the identity, the more terms we need in
the product $\prod_i^n T_i=T$. Hence, $n$ will be an increasing
function of $\epsilon$ and the question whether or not one can
safely replace each $T_i$ by a Markovian channel amounts to the
estimation of an accumulated error of the form ``$n\epsilon$''.
The following theorem shows that the scaling of the latter is
benign so that indeed $\overline{\cal I}=\overline{{\cal I}'}$.
Moreover, since $\overline{{\cal I}'}\subseteq\overline{{\cal
J}}\subseteq\overline{{\cal I}}$ both sets are equal to the set of
continuous completely positive evolutions.
\begin{theorem}[Structure of infinitesimal divisible channels]\label{thm:structure}
With the above notation we have that $\overline{\cal
I}=\overline{{\cal I}'}=\overline{{\cal J}}$. In particular, every
infinitesimal divisible channel can be arbitrary well approximated
by a product of Markovian channels.
\end{theorem}
\begin{proof}
We want to show that one can replace every channel $T_i$ in the
decomposition $T=\prod_{i=1}^n T_i$ with $||T_i-\id||\leq\epsilon$
by a Markovian channel such that the error becomes negligible in
the limit $\epsilon\ra 0$ (and thus $n\rightarrow\infty$). This is
proven in two steps: (i) we calculate the error obtained from the
Markovian approximation in Lem.\ref{lemma:Mapprox} as a function
of $n$ and $\epsilon$, and (ii) we relate $n$ and $\epsilon$ by
exploiting properties of the determinant shown in
Thms.\ref{thm:detMarkov},\ref{thm:detBounds}. Strictly speaking,
we will in both steps not use the distance $\epsilon$ to the
identity but rather a distance $\delta\leq\epsilon$ to a nearby
unitary.

First we write $\prod_iT_i =\prod_i\tilde{T}_i U_i$ where
$\tilde{T}_i=T_iU_i^{-1}$ is such that $(\tilde{T}_i-\id)$ is a
purely dissipative generator according to Lem.\ref{lemma:Mapprox}.
The idea is then to approximate $\tilde{T}_i$ by
$\exp(\tilde{T}_i-\id)$. The total error in this approximation is
then given by \be\label{tzu} \Big|\Big|\prod_i T_i - \prod_i
e^{\tilde{T}_i-\id}U_i\Big|\Big| = \Big|\Big| \prod_i T_i-
\prod_i(T_i+\Delta_i)\Big|\Big|\;,\ee where
$\Delta_i=(\exp[\tilde{T}_i-\id]-\tilde{T}_i)U_i$ is an operator
whose norm vanishes as $O(||\tilde{T}_i-\id||^2)$. The product
$\prod_i(T_i+\Delta_i)$ contains $({n\atop{k}})$ terms of the form
``$T_i^{n-k}\Delta_i^k$'' where the $T_i$s come in at most $k+1$
groups for each of which we can bound the norm by $\sqrt{d}$
\cite{PWPR06}. If we define $\delta:=\max_i ||\tilde{T}_i-\id||$
we can therefore bound the error in Eq.(\ref{tzu}) by \be
\Big|\Big|\prod_i T_i-\prod_i (T_i+\Delta_i)\Big|\Big|\leq
\sqrt{d}\Big(\big(1+\sqrt{d}O(\delta^2)\big)^n-1\Big)\;.\ee This
vanishes iff $\delta^2 n \ra 0$ as $n\ra\infty$.\footnote{An
alternative way for obtaining this result is by defining
$C(l):=\prod_{i=1}^l T_i\prod_{j=l+1}^n (T_j+\Delta_j)$. Then
Eq.(\ref{tzu}) equals $||C(n)-C(0)||=||\sum_{k=0}^{n-1}
C(k+1)-C(k)||\leq (n+1)d\delta^2$ where the inequality follows
from the triangle inequality.}

To relate $\delta$ and $n$ we use Thm.\ref{thm:detMarkov} from
which we obtain $\delta\leq -\frac2d\min_i\ln\det
\exp(\tilde{T}_i-\id)$. Exploiting continuity of the
determinant\footnote{$|\det A-\det B|\leq d ||A-B||
\max\{||A||,||B||\}^{d-1}$ \cite{Bhatia}} and denoting by
$\tilde{T}_\delta$ the channel $T_i$ giving rise to the maximum
distance $\delta$, this gives \be \delta\leq\label{eq:detlog}
-\frac2d\ln\left[\det\tilde{T}_\delta-O(\delta^2)\right].\ee Since
by assumption there are arbitrarily fine-grained decompositions
$T=\prod_iT_i$ we can w.l.o.g. assume that all $T_i$ have equal
determinant $\det T_i=\big(\det T\big)^{1/n}$ (or ones distributed
within a sufficiently narrow interval). As $\det \tilde{T}_i=\det
T_i$ Eq.(\ref{eq:detlog}) relates $n$ and $\delta$---unfortunately
in a way that we cannot yet conclude that $\delta=o(n^{-1/2})$.
However, it enables us to lift any polynomial bound to higher
order: assume that $\delta=O(n^{-q})$ for some $q\in(0,1)$. Then
Eq.(\ref{eq:detlog}) gives rise to $\delta=O\big(n^{-2q}-(\ln \det
T)/n\big)$ which leads recursively to $\delta=O(1/n)$ provided
that $\det T>0$. Hence, any bound of the form $\delta=O(n^{-q})$,
$q>0$ will suffice to show that the error given by Eq.(\ref{tzu})
vanishes asymptotically. Such a bound is provided by
Thm.\ref{thm:detBounds}as we obtain from Eq.(\ref{scaling}) that
$\delta=O\big(\sqrt{-(\ln\det T)/n} \big)$.

Note finally that it suffices to consider the case $\det T> 0$ as
singular channels are only included in Def.\ref{Def:divisibility}
by taking the closure of ${\cal I}$ and $\det T<0$ is excluded by
Prop.\ref{prop:detpos}.
 \qed
\end{proof}

 Similar to the notion of divisibility we may introduce
\emph{infinitesimal divisible positive maps} by replacing $\tf^+$
in Def.\ref{def:infinitesimal} by $\tf$. In both cases we can
again decide whether a map is infinitesimal divisible by
considering its normal form under invertible filtering operations
with Kraus rank one:
\begin{theorem}[Reduction to normal form]\label{Thm:IDSLOCC}
Let ${\cal T}\in\{\tf,\tf^+\}$ and $T,\tilde{T}\in{\cal T}$ be
related via $T= T_A\tilde{T}T_B$ where $T_A,T_B\in\p^+$ are
invertible completely positive maps with Kraus rank one. Then $T$
is infinitesimal divisible iff $\tilde T$ is.
\end{theorem}
\begin{proof}
As the statement is symmetric in $T$ and $\tilde{T}$ (due to
invertibility of $T_A,T_B$) it is sufficient to prove one
direction. So let us assume that
$\tilde{T}=\prod_{i=1}^n\tilde{T}_i$ is infinitesimal divisible.
Then we can write $T=\prod_{i=1}^n R_i\tilde{T}_i R_{i+1}^{-1}$
where $R_i\in\p^+$ are invertible maps of Kraus rank one with
$R_1=T_A$ and $R_{n+1}^{-1}=T_B$. We will now show that the
intermediate $R_i$'s can be chosen such that $T_i:=R_i\tilde{T}_i
R_{i+1}^{-1}\in{\cal T}$ is such that $||T_i-\id||$ vanishes
uniformly as $||\tilde{T}_i-\id||\leq\epsilon\ra 0$. This is
achieved by recursively constructing $R_{i+1}$ from $R_i$
according to the proof of Thm.\ref{thm:nontp} and exploiting that
\be ||T_i-\id||\leq ||R_iR_{i+1}^{-1}-id||+\epsilon
||R_i||\;||R_{i+1}^{-1}||.\label{eq:Tiepsilonbound}\ee Let us
denote by $K_i=U_iP_i$ the polar decomposition of the Kraus
operator of $R_i(\cdot)=K_i\cdot K_i^\dagger$. The trace
preserving requirement for $T_i$ imposes that $R_{i+1}^{-1*}
\tilde{T}_i^*R_i^*(\1)=\1$ which is achieved by choosing
$P_{i+1}=\sqrt{\tilde{T}_i^*(P_i^2)}$. As any unital positive map
is spectrum-width decreasing \cite{POVMs} we have for the range of
eigenvalues
$\big[\lambda_{min}(P_{i+1}),\lambda_{max}(P_{i+1})\big]\subseteq
\big[\lambda_{min}(P_{i}),\lambda_{max}(P_{i})\big]$. This allows
us to bound the second term in Eq.(\ref{eq:Tiepsilonbound}) by
$\epsilon ||R_i||\;||R_{i+1}^{-1}||\leq\epsilon
\lambda_{max}^2(P_1)\lambda_{min}^{-2}(P_1)$.

To bound the first term note that
$||\tilde{T}_i^*(P_i^2)-P_i^2||\leq\epsilon
||P_i^2||_2\leq\epsilon \lambda_{max}^2(P_i)\sqrt{d}$. By
continuity of the square root\footnote{$||\sqrt{A}-\sqrt{B}||\leq
||A-B||^{1/2}$ for all positive $A,B$ \cite{Bhatia}.} this implies
$||P_{i+1}-P_i||\leq \sqrt{\epsilon}d^{1/4}\lambda_{max}(P_i)$.
Hence, $||P_iP_{i+1}^{-1}-\1||\leq
\sqrt{\epsilon}d^{1/4}\lambda_{max}(P_i)\lambda_{min}^{-1}(P_{i+1})$
yielding a $\sqrt{\epsilon}$ bound for the first term in
Eq.(\ref{eq:Tiepsilonbound}) if we take $U_{i+1}=U_i$. The latter
choice might not be possible in the $n$'th step (as the trace
preserving requirement only fixes $T_B$ up to a unitary
conjugation). However, we can always add an additional unitary
without changing the property of being infinitesimal divisible.
\qed\end{proof}

Note that the above reduction to normal form together with
Thm.\ref{thm:structure} preserves continuity in the sense that if
$T=T_A e^L T_B$ with Markovian $e^L\in\tf^+$, then we can write\be
T=\mathbb{T}\;e^{\int_0^1L(\tau)d\tau}\;,\ee where $\mathbb{T}$ is
the time-ordering operator and $\tau\mapsto L(\tau)$ is a
continuous mapping onto generators of the form in
Eqs.(\ref{normal1},\ref{normal2}). In other words, $T$ is then a
solution of a time-dependent master equation $d\rho/dt=L(t)\rho$.
The fact that every generic infinitesimal divisible channel can be
written in this way is proven below for the case $d=2$ of qubit
channels.

\section{Qubit Channels}\label{sec:qubit}

The simplicity of qubit channels ($T:\M_2\ra\M_2$) often allows a
more thorough analysis of their properties. An exhaustive
investigation of the convex structure of the set of qubit channels
and positive trace-preserving qubit maps was for instance given in
\cite{2convexcp} and \cite{2convex} respectively. Similarly, their
normal form under invertible filtering operations was determined
in \cite{filtering}. In the following we will make extensive use
of these results in order to derive a complete characterization of
the above discussed semigroup structure of this set. We begin by
recalling some of the basic tools and treat the case of extremal
qubit channels (two Kraus operators) first, as later argumentation
will build up on this. The main results---a complete
characterization of divisible and infinitesimal divisible qubit
channels---are then stated in Thm.\ref{thm:ind2} and
Thm.\ref{thm:invv2}.

The representation we will mainly use in the following is  a real
$4\times4$ matrix $\hat{T}_{ij}:=\tr[\sigma_iT(\sigma_j)]/2$ (cf.
\cite{2convexcp}) which is in turn characterized by a $3\times 3$
block $\Delta$ and a vector $v\in\mathbb{R}^3$  encoding the
correlations and the reduced density matrix of the Jamiolkowski
state respectively:
\be\label{T2b} \hat{T}=\left(%
\begin{array}{cc}
  1 & 0\\
  v & \Delta \\
\end{array}%
\right).\ee
 Since there
is an epimorphism from $SU(2)$ to the rotation group $SU(3)$ we
can always diagonalize $\Delta$ by acting unitarily before and
after $T$. More specifically, for any $T\in\tf$ there exist
unitary conjugations $U_1,U_2$ such that $U_1TU_2$ has
$\Delta={\rm diag}(\lambda_1,\lambda_2,\lambda_3)$ with
$1\geq\lambda_1\geq\lambda_2\geq|\lambda_3|$. Expressing complete
positivity in terms of $v$ and $\lambda$ is rather involved and
discussed in detail in
\cite{unitalqubits,unitalqubits2,2convexcp}. A necessary condition
for complete positivity is that \be
\lambda_1+\lambda_2\leq1+\lambda_3\;,\label{eq:cpcond}\ee which
becomes sufficient if the channel is unital, i.e., $v=0$.

A very useful standard form for qubit channels is obtained when
building equivalence classes under filtering operations
\cite{filtering}.\footnote{This standard form is referred to as
\emph{Lorentz normal form} as the mapping $T\mapsto T_AT T_B$
corresponds to $\hat{T}\mapsto L_A\hat{T}L_B$ where $L_{A,B}$ are
proper orthochronous Lorentz transformations \cite{filtering}.}

\begin{theorem}[Lorentz normal form]\label{Thm:SLOCC} For every qubit channel $T\in\tf^+$ there exist invertible
$T_A,T_B\in\p^+$, both of Kraus rank one, such that $T_A T
T_B=\tilde{T}\in\tf^+$ is of one of the following three
forms:\begin{enumerate}
    \item \emph{Diagonal}: $\tilde{T}$ is unital ($v=0$). This is
    the generic case.
    \item \emph{Non-diagonal}: $\tilde{T}$ has $\Delta={\rm
    diag}(x/\sqrt{3},x/\sqrt{3},1/3)$, $0\leq x\leq 1$ and
    $v=(0,0,2/3)$. These channels have Kraus rank 3 for $x<1$ and
    Kraus rank 2 for $x=1$.
    \item \emph{Singular}: $\tilde{T}$ has $\Delta=0$ and
    $v=(0,0,1)$. This channel has Kraus rank 2 and is singular in
    the sense that it maps everything onto the same  output.
\end{enumerate}
\end{theorem}
A concatenation of qubit channels $T_1T_2=T$ corresponds to a
multiplication of the respective matrices
$\hat{T}_1\hat{T}_2=\hat{T}$ so that $\Delta_1\Delta_2=\Delta$ and
$\Delta_1v_2+v_1=v$. In this way we can for instance decompose
every channel of the second form  in Thm.\ref{Thm:SLOCC} into
\be\label{eq:decomp2}
{\footnotesize\left(%
\begin{array}{cccc}
  1 &  &  &  \\
   & x/\sqrt{3} &  &  \\
   &  & x/\sqrt{3} &  \\
  2/3 &  &  & 1/3 \\
\end{array}%
\right)=\left(%
\begin{array}{cccc}
  1 &  &  &  \\
   & 1/\sqrt{3} &  &  \\
   &  & 1/\sqrt{3} &  \\
  2/3 &  &  & 1/3 \\
\end{array}%
\right)\left(%
\begin{array}{cccc}
  1 &  &  &  \\
   & x &  &  \\
   &  & x &  \\
   &  &  & 1 \\
\end{array}%
\right)},\ee which is a concatenation of two Kraus rank-two
channels (unless $x=1$ where the initial channel is already
rank-two). Let us now have a closer look at qubit channels with
Kraus rank two.

\subsection{\bf Extremal qubit channels}

Channels with Kraus rank two play  an important role regarding the
convex structure of the set of qubit channels. It was shown in
\cite{2convexcp} that every extreme point of this set is either a
unitary conjugation or a (non-unital) Kraus rank-two channel. In
this context it has been shown that every Kraus rank-two channel
can up to unitary conjugations be represented by\be\label{eq:2uv} \hat{T}={\footnotesize{\left(%
\begin{array}{cccc}
  1 &  &  &  \\
   & c_u &  & \\
   &  & c_v &  \\
  s_u s_v &  & & c_u c_v \\
\end{array}%
\right)}},\quad c_u=\cos u,\ s_u=\sin u\,.\ee For the remainder of
this subsection we will, however, use a different representation
which is very handy for our purposes albeit less explicit than the
one in Eq.(\ref{eq:2uv}). This will allow us to prove the
following:

\begin{theorem}[Infinitesimal divisibility of Kraus rank-two
channels]\label{Thm:rank2} Let $T:\M_2\ra\M_2$ be a qubit channel
with Kraus rank two. Then there exist unitary conjugations $U_1,
U_2$, a continuous time-dependent Lindblad generator $L$ and $t>0$
such that
\begin{equation}\label{CAMark}
 U_1TU_2 = \mathbb{T}\; e^{\int_0^t L(\tau) d\tau}\;.
\end{equation}
\end{theorem}
In order to prove this result, we will first introduce the
mentioned normal form and then explicitly construct the Lindblad
generators. To this end consider the set of specific channels
$C\in\tf^+$ with Kraus operators
\begin{subequations}
\label{Aes}
\begin{eqnarray}
 A_1 &=& |0\rangle\langle a|,\\
 A_2 &=& |0\rangle\langle b| + x |1\rangle\langle 1|.
\end{eqnarray}
\end{subequations}
We will take $x$ and the zero components of $|a\rangle$ and
$|b\rangle$ real. The trace preserving condition gives
\begin{equation}
\label{TP}
 |a\rangle\langle a| + |b\rangle\langle b| = \1- x^2 |1\rangle\langle
 1|.
 \end{equation}
We will prove that all channels $C$ are of the form on the r.h.s.
of Eq.(\ref{CAMark}), which, together with the following Lemma,
will yield the proof of the theorem.

\begin{lemma}
 For any qubit channel $T$ with Kraus rank two, there exist unitary
 conjugations
 $U_1,U_2$ such that $T=U_1CU_2$.
\end{lemma}
\begin{proof}
Given the Kraus operators $K_{1,2}$ of $T$, we can always find
$\alpha_{1,2}$ such that $\alpha_1  K_1 + \alpha_2  K_2$ has rank
1 (i.e., zero determinant). Thus, a different set of Kraus
operators can be chosen with $\hat K_1 = |e_0\rangle\langle f_1|$,
and $\hat K_2 = |e_0\rangle\langle f_2| + |e_1\rangle\langle
f_3|$, where $e_{0,1}$ are orthonormal. Defining $A_i=V_1\hat K_i
V_2$, with $V_1,V_2$ unitaries, using the fact that we can
multiply Kraus operators with complex numbers of unit modulus, and
imposing that the channel is trace preserving, we easily reach the
above form.\qed
\end{proof}

Thus, from now on we concentrate on the specific channels $C$.
Depending on the vectors $a,b$, we can have very different
channels. We define:
\begin{definition} Given a channel of the above form $C$, we will call
it: (i) class-1 if $\langle a|0\rangle=\langle b|1\rangle=0$; (ii)
class-2 if it is not in class-1 and $x=1$; (iii) class-3
otherwise.
\end{definition}

The main difference between these channels lies on the number of
pure states that are mapped into pure states. In fact, it can be
easily checked that for all channels $|0\rangle\to |0\rangle$ and
that for class-1 channels, either all pure states are mapped into
$|0\rangle$ (for $x=0$) or only $|0\rangle$ is mapped into a pure
state (for $x\ne 0$), whereas for class-2 and 3, apart from
$|0\rangle$, there is only one state $|c\rangle \perp |a\rangle$
which is mapped into a pure state. In the following we will
consider the different classes of channels independently.
\paragraph{Class-1 channels}
We can write $|a\rangle=(1-x^2)^{1/2}|1\rangle$ and
$|b\rangle=|0\rangle$, so that all these channels are parametrized
just by $x$, and therefore we will write $C_x$. We have
\begin{equation}
\label{semiGC}
 C_{x_1}C_{x_2}=C_{x_1x_2}, \quad C_1=1.
\end{equation}
Thus, this class forms a continuous 1-parameter semigroup. Using
infinitesimal transformations one can easily show that
\begin{equation}
\label{MarkovClass1}
 C_x= e^{-\ln(x) L},\quad L(\rho)=2|0\rangle\langle 1|\rho|1\rangle\langle 0|-
\rho|1\rangle\langle 1|-|1\rangle\langle 1|\rho.
\end{equation}
\paragraph{Class-2 channels}
In this case we can write $|a\rangle=(1-y^2)^{1/2}|0\rangle$ and
$|b\rangle=y|0\rangle$, so that again we have a single parameter
family $C_y$. As before, we obtain a one-parameter semigroup
$C_y=\exp(-\ln(y)L)$ but now with $L(\rho)=2\sigma_z\rho\sigma_z-
2\rho$.
\paragraph{Class-3 channels}
We show now that every channel $C$ in this class is completely
determined by the vector different from $|0\rangle$ which is
mapped into a pure state.

As mentioned above, this class is characterized by the fact that a
normalized pure state $|c\rangle \perp |a\rangle$ is mapped into
another pure $|c'\rangle$:
\begin{subequations}
\label{ccprime} \bea
 |c\rangle&=&c_0 e^{i\varphi} |0\rangle + c_1|1\rangle,\quad
c_0,c_1\in\mathbb{R}\\
 |c'\rangle &=& y c_0 e^{i\varphi} |0\rangle + x c_1|1\rangle,
\eea
\end{subequations}
where $y\ge 1$ ensures  normalization. That is, since $x<1$, the
distance to the vector $|0\rangle$ decreases, whereas the azimutal
angle in the Bloch sphere remains constant. Now we will show the
converse:
\begin{lemma}
 Given $|c\rangle$ and $|c'\rangle$ as in Eq.(\ref{ccprime}) with $x<1$, there
 exists a unique class-3 channel which maps
 $|c\rangle\to|c'\rangle$.
\end{lemma}
\begin{proof}
The definition of $c$ and $c'$ fixes the values of $x$ and
$|a\rangle$ up to a normalization for the Kraus operators
(\ref{Aes}). Both $||a||$ and $|b\rangle$ are completely specified
by the condition (\ref{TP}). Indeed, defining $|\tilde
a\rangle:=|a\rangle/||a||$ we have to fulfill that
$\1-x^2|1\rangle\langle 1|-||a||^2 |\tilde a\rangle\langle \tilde
a|=|b\rangle\langle b|$, i.e., has rank 1, which automatically
fixes
\begin{equation}
 ||a||^2=\frac{1-x^2}{1-x^2c_1^2}
\end{equation}
and thereby $|b\rangle$ through Eq.(\ref{TP}).\qed
\end{proof}

The maps in this class are parametrized by $x,c_1\in (0,1)$ and
$\varphi\in [0,2\pi)$, and thus we will write $C_{c_1,x,\varphi}$.
They fulfill
\begin{equation}
 C_{xc_1,y,\varphi}C_{c_1,x,\varphi}=C_{c_1,xy,\varphi}.
\end{equation}
Note that $C_{c_1,x,\varphi}\to \id$ for $x\to 1$. Thus, we can
determine the generator of an infinitesimal transformation as $
 L_{c_1,\varphi}:=\lim_{\epsilon\to
 0}{(\id-C_{c_1,e^{-\epsilon},\varphi})}/{\epsilon}.
$ We obtain \bea
 L_{c_1,\varphi}(\rho)&=&i[\rho,H_{c_1,\varphi}] +
 D_{c_1,\varphi}(\rho),\\
 H_{c_1,\varphi} &=& \frac{c_1}{i c_0} \big(e^{i\varphi}|0\rangle\langle 1|-
 e^{-i\varphi}|1\rangle\langle 0|\big),
\eea and $D_{c_1,\varphi}$ is a simple dissipative Lindblad
generator characterized by a single Kraus operator of the form $
 A_{c_1,\varphi}= {\sqrt{2}}|0\rangle(c_1\langle 0|-c_0\langle
 1|)/{c_0} ,
$ with $c_0=(1-c_1^2)^{1/2}$. Thus, we arrive at the result of
Thm.\ref{Thm:rank2} and can write
\begin{equation}
 C_{c_1,x,\varphi}= \mathbb{T}\; \exp{\int_0^{-\ln(x)}\!
 L_{c_1e^{-\tau},\varphi}d\tau}.
\end{equation}

\subsection{\bf Divisible and indivisible qubit channels}

We are now prepared to give a complete characterization of
divisible/indivisible qubit channels. An indivisible example---the
channel with minimal determinant---was already given in
 Corollary \ref{Cor:IndCP}.
Surprisingly, there are indivisible channels with positive
determinant as well:

\begin{theorem}[Indivisible qubit channels]\label{thm:ind2}
A non-unitary qubit channel is indivisible within $\tf^+$ if and
only if it has Kraus rank three and its Lorentz normal form
(Thm.\ref{Thm:SLOCC}) is diagonal (i.e., unital).
\end{theorem}
\begin{proof}
As all qubit channels with Kraus rank four are divisible due to
Thm.\ref{Thm:GenericDivisibility} and all rank-two channels are
divisible according to the previous subsection, the Kraus rank of
indivisible qubit channels must be three (or one---trivially).
Following Cor.\ref{Cor:NormalformReduction} it suffices to
consider the Lorentz normal form of Thm.\ref{Thm:SLOCC}. Since the
non-diagonal case can be decomposed via Eq.(\ref{eq:decomp2}) into
\emph{divisible} Kraus rank-two channels, it remains to show that
all unital channels with Kraus rank three are indivisible.

Suppose $T$ is such a channel and we can write $T={T}_1'{T}_2'$
with non-unitary ${T}_i'\in\tf^+$. Then there is also a
decomposition $T=T_1T_2$ into non-unitary \emph{unital} channels
$T_i$ which can for instance be obtained by setting the $v$'s in
$\hat{T}_i'$ in Eq.(\ref{T2b}) to zero and keeping
$\Delta_i=\Delta_i'$. This will still be a decomposition of $T$
but neither change the determinant (and thus non-unitarity) nor
complete positivity as Eq.(\ref{eq:cpcond}) becomes a necessary
and sufficient condition for unital channels.

By assumption the Jamiolkowski state
$\tau=(T_1T_2\otimes\id)(\omega)$ has rank three. As unital qubit
channels are convex combinations of unitary conjugations and
$(U\otimes\id)(\omega)=(\id\otimes U^T)(\omega)$ we can write
\be\tau=(T_1\otimes T_2^T)(\omega)\;,\label{eq:tauT}\ee where
$T_2^T$ is again a unital
 channel whose Kraus operators are related to those of $T_2$ by
transposition. It follows from Eq.(\ref{eq:tauT}) that the Kraus
rank of $T_1$ and $T_2$ is at most three. Assume now $T_2$ has
Kraus rank three. Then $\tau \geq\mu(T_1\otimes\id)(\1-\Omega)$
where $\Omega$ is the projector onto a maximally entangled state
and $\mu$ is the smallest non-zero eigenvalue of the Jamiolkowski
state of $T_2$. Thus, if $\{\sqrt{p_i}U_i\}$ are the Kraus
operators of $T_1$ with $\{U_i\}$ orthogonal unitaries and
$\{p_i\}$ probabilities, then \be \tau \geq \mu \Big(\1-\sum_{i}
p_i (U_i\otimes\1)\Omega(U_i\otimes\1)^\dagger\Big)\;.\ee Since
the projectors in the sum are orthogonal, $\tau$ can only be rank
deficient if there is only a single term in the sum and $T_1$ thus
a unitary.

The only remaining possibility is thus a decomposition into two
unital channels each of Kraus rank two. In order to rule this out
note that in this case the support of $\tau$ equals that of
\be\label{eq:suptau} \1-\sum_{i=1}^2 p_i
(U_i\otimes\1)P(U_i\otimes\1)^\dagger,\ee where $P$ is now some
two-dimensional projector. Denoting by $\psi$ the normalized and
maximally entangled null vector of $\tau$ we have to have that
$P(U_i\otimes\1)^\dagger|\psi\rangle=(U_i\otimes\1)^\dagger|\psi\rangle$
so that \be P=\sum_{j=1}^2
(U_j\otimes\1)^\dagger|\psi\x\psi|(U_j\otimes\1).\ee Now we
exploit the fact that every basis  of orthogonal unitaries
$\{U_j\}$ in $\M_2$ is essentially equivalent to the Pauli basis
in the sense that there are always unitaries $V_1,V_2$ and phases
$e^{i\varphi_j}$ such that $U_j=V_1\sigma_j V_2 e^{i\varphi_j}$
\cite{2special}. It follows that $U_1U_2^\dagger$ equals
$U_2U_1^\dagger$ up to a phase which in turn implies that the
expression in Eq.(\ref{eq:suptau}) and thus $\tau$ have rank
two---contradicting the assumption and therefore concluding the
proof.\qed
\end{proof}

\subsection{\bf Infinitesimal divisible qubit channels}

We will now give a necessary and sufficient criterion for qubit
channels to be infinitesimal divisible, formulated in terms of the
matrix representation Eq.(\ref{T2b}) of the channel's Lorentz
normal form (Thm.\ref{Thm:SLOCC}):
\begin{theorem}[Characterization of infinitesimal divisible
channels]\label{thm:invv2} Consider a qubit channel and denote by
$s_{min}$ the smallest singular value of the $\Delta$-block of its
Lorentz normal form. The channel is infinitesimal divisible iff
one of the following conditions is true\begin{enumerate}
    \item The Lorentz normal form is not diagonal.
    \item The normal form is diagonal and ${\rm rank}(\Delta)<2$.
    \item The normal form is diagonal and \be\label{eq:condID2}
s^2_{min}\geq \det \Delta > 0\;.\ee
\end{enumerate}
\end{theorem}
\begin{proof}We exploit the fact that by Thm.\ref{Thm:IDSLOCC} a channel is infinitesimal divisible iff its Lorentz normal form is.
If the  normal form is not diagonal, then by
Eqs.(\ref{eq:decomp2},\ref{eq:2uv}) it has Kraus rank two or is a
product of Kraus rank-two channels which are in turn infinitesimal
divisible according to Thm.\ref{Thm:rank2}. Similarly, if the
normal form is diagonal and $\Delta={\rm diag}(\lambda,0,0)$ we
can again factorize it into Kraus rank-two channels as
$\Delta={\rm diag}(1,0,0){\rm diag}(\lambda,1,\lambda)$. To
complete point 2. in the theorem note that the unital channel with
$\Delta=0$ is a limit of a Markovian unital channel as
$\Delta=\lim_{t\ra\infty} e^{-t}\1$.

Consider now the generic case where the Lorentz normal form is
diagonal and $\det T\neq 0$. Following Prop.\ref{prop:detpos} we
have that $\det T\geq 0$ for every infinitesimal divisible
channel. Moreover, by Thm.\ref{thm:structure} we can express these
channels in terms of products of Markovian channels, which can
w.l.o.g. be chosen unital. The latter can in turn be decomposed
into even simpler pieces by exploiting the Lie-Trotter formula
$\lim_{n\ra\infty}\big(e^{L_1/n}e^{L_2/n}\big)^n=e^{L_1+L_2}$. In
this way every unital Markovian qubit channel can be written as a
product of unitaries and unital Kraus rank-two channels with
$\Delta={\rm diag}(1,\lambda,\lambda)$ \cite{Bacon}. Note that for
these channels we have $s_{min}^2=\det T$. The inequality
Eq.(\ref{eq:condID2}) follows then from concatenating these
channels together with multiplicativity of the determinant and the
fact that $s_{min}(\Delta_1)s_{min}(\Delta_2)\leq
s_{min}(\Delta_1\Delta_2)$.

Let us now show the converse, i.e., that Eq.(\ref{eq:condID2})
together with a diagonal Lorentz normal form implies that the
channel is infinitesimal divisible. To this end we introduce
$\Delta_t:=\exp{(t\ln\Delta)}$, $t\geq0$ and show that it
corresponds to a completely positive unital semigroup if $\Delta$
(chosen positive definite and diagonal) satisfies
Eq.(\ref{eq:condID2}). Following Eq.(\ref{eq:cpcond}) we have to
show that $\tr{\Delta_t}\leq 1+2s_{min}(\Delta_t)$ for complete
positivity. Moreover, it suffices to prove this for infinitesimal
$t$ since larger times are obtained by concatenation which
preserves complete positivity. In leading order we get \bea
\tr{\Delta_t}&=& \tr{\big[\1+t\ln\Delta\big]}+O(t^2)\\
&=& 3+t\ln\det T+O(t^2)\;,\\
1+2s_{min}(\Delta_t)&=&1+2\big(1+t\ln
s_{min}(\Delta)\big)+O(t^2)\;, \eea from which we obtain \be
\tr{\Delta_t}-\big[1+2s_{min}(\Delta_t)\big]=t\big[\ln\det(\Delta)-\ln
s_{min}^2(\Delta)\big]\pm O(t^2)\;,\ee which is indeed negative
for infinitesimal $t$ if $\det(\Delta)<s_{min}^2(\Delta)$. The
case of equality is covered by the fact that we can then express
$\Delta={\rm diag}(\lambda_1,\lambda_2,\lambda_1\lambda_2)={\rm
diag}(\lambda_1,1,\lambda_1){\rm diag}(1,\lambda_2,\lambda_2)$ as
concatenation of two Kraus rank-two channels.

What remains to discuss is the case of a diagonal normal form with
$\Delta={\rm diag}(\lambda_1,\lambda_2,0)$, $\lambda_i>0$. Note
that channels of zero determinant can be infinitesimal divisible
due to the fact that we took the closure in
Def.\ref{def:infinitesimal}. Hence, there must be an infinitesimal
divisible channel $T_\epsilon$ with non-zero determinant in every
$\epsilon$-neighborhood of $T$. If $T$ is unital we can again
w.l.o.g. chose $T_\epsilon$ to be unital as well. In leading order
the $\Delta$-block of $T_\epsilon$ has singular values
$\lambda_1,\lambda_2$ and $\epsilon$. For sufficiently small
$\epsilon$ this can, however, never satisfy Eq.(\ref{eq:condID2})
 so that there cannot be an infinitesimal divisible channel with non-zero determinant close to $T$ and thus $T$ itself
 cannot be infinitesimal divisible.\qed
\end{proof}

Thm.\ref{thm:invv2} characterizes the set of qubit channels which
are solutions of continuous time-dependent master equations for
completely positive evolutions. As in the theory of open quantum
systems complete positivity is often dropped in the context of
time-dependent master equations we provide the analogous statement
for evolutions which are (locally) merely positivity preserving:
\begin{theorem}[Continuous positive evolutions]
A qubit channel $T\in{\tf}^+$ is infinitesimal divisible within
the set $\tf$ of positive trace preserving maps iff it has
non-negative determinant.
\end{theorem}
\begin{proof}
By multiplicativity and continuity of the determinant we know that
$\det T\geq 0$ is indeed necessary for $T$ to be infinitesimal
divisible. In order to prove sufficiency we exploit once again the
Lorentz normal form together with Thm.\ref{Thm:IDSLOCC} and the
fact that the sign of the determinant does not change upon
concatenating with Kraus rank-one filtering operations. If the
normal form is not diagonal, then $T$ is infinitesimal divisible
according to Thm.\ref{thm:invv2}. If the normal form is diagonal
and $\det T>0$, then the statement follows from the fact that the
corresponding unital channel is an element of a positivity
preserving semigroup given by $\Delta_t=\exp[t\ln \Delta]$. As
$\Delta_t\leq\1$ for all $t\geq 0$ the corresponding map is always
positive. The remaining cases with $\det T=0$ are obtained by
taking the closure.\qed
\end{proof}

\section{Conclusion}

We have mainly addressed two questions: which quantum channels can
be broken down into infinitesimal pieces, and which can be
expressed as a non-trivial concatenation of other channels at all.
This led us to the two notions of \emph{infinitesimal
divisibility} and \emph{divisibility} respectively. Loosely
speaking, the former class corresponds to the set of solutions of
time-dependent master equations. However, to make this a strong
correspondence  continuity of the Liouville operator (at least
piecewise) would clearly  be desirable. This follows from our
analysis only for qubit channels for which a rather exhaustive
characterization was possible. For higher dimensions a similar
complete classification might be hard to obtain unless one
restricts to specific classes like diagonal or quasi-free channels
\cite{quasifree}.

We find it remarkable that in the vicinity of the ideal channel
all types of channels can be found (i.e., indivisible, divisible,
not infinitesimal divisible, Markovian, etc.). This is, in fact,
what makes the proof of our main structure theorem
non-trivial---if all channels close to the identity would be
Markovian, it would follow immediately.

Apart from the implications for the theory of open quantum systems
and the abstract semigroup structure of the set of quantum
channels we can think of applying the techniques and results
presented in this work in various contexts.

Renormalization-group transformations for quantum states on a spin
chain \cite{RG} for instance use concatenations and---in the
infrared limit---divisions of quantum channels.

Moreover, when considering quantum channels with a classical
output in the sense of the positive operator valued measure (POVM)
formalism, then a similar train of thoughts leads to the notion of
\emph{clean} POVMs which cannot be expressed as a non-trivial
concatenation of a quantum channel with a different POVM
\cite{POVMs}.

Finally, it would be interesting to know whether a concatenation
of quantum channels allows for a quantitative estimate  of the
channel capacity based on the capacities of the constituents which
goes beyond the trivial bottleneck-inequality. In this context
also the stability of the above introduced notions under tensor
products is an interesting problem.

\begin{acknowledgements} We thank T. Cubitt, J. Eisert and A. Holevo for
valuable discussions.
\end{acknowledgements}


\begin{thebibliography}{30}

\bibitem{Holevo} A.S. Holevo, {\it Statistical Structure of
Quantum Theory}, Springer Lecture Notes in Physics (2001).
\bibitem{Horn} R.A. Horn, Z. Wahrscheinlichkeitstheorie und Verw.
Gebiete {\bf 8}, 219 (1967).
\bibitem{Hinf} A.S. Holevo, Theor. Probab. Appl. {\bf 32}, 560
(1986).
\bibitem{Denisov} L.V. Denisov, Th. Prob. Appl. {\bf 33}, 392
(1988).
\bibitem{Jami} A. Jamiolkowski, Rep. Math. Phys. {\bf 3} 275 (1972).
\bibitem{Choi} M.D. Choi, Lin. Alg. Appl. {\bf 10}, 285 (1975).
\bibitem{Kraus} K. Kraus, {\it States, Effects, and Operations},
Springer (1983).
\bibitem{WPG} M.M. Wolf, D. Perez-Garcia, quant-ph/0607070 (2006).
\bibitem{Lindblad} G. Lindbald, Commun. Math. Phys. {\bf 48}, 119
(1976).
\bibitem{Gorini} V. Gorini, A. Kossakowski, E.C.G. Sudarshan, J.
Math. Phys. {\bf 17}, 821 (1976).
\bibitem{snapshot}  M.M. Wolf, J. Eisert, T.S. Cubitt, J.I. Cirac,
arXiv:0711.3172 (2007).
\bibitem{Davies} E.B. Davies, Rep. Math. Phys. {\bf 17}, 249 (1980).
\bibitem{PWPR06} D. Perez-Garcia, M.M. Wolf, D. Petz, M.B. Ruskai,
J. Math. Phys. {\bf 47}, 083506 (2006).
\bibitem{Dio} W.M. Schmidt, {\it Diophantine Approximation}, Lecture Notes in Math.
785, Springer Verlag, 1980.
\bibitem{Bhatia} R. Bhatia, {\it Matrix Analysis}, Springer
Graduate Texts in Mathematics 169 (1997).
\bibitem{Streater} R.F. Streater, {\it Statistical Dynamics},
Imperial College Press (1995).
\bibitem{Wigner} E.P. Wigner, {\it Gruppentheorie}, Vieweg (1931);
{\it Group Theory}, Academic Press (1959).
\bibitem{Bargmann} V. Bargmann, J. Math. Phys. {\bf 5}, 862 (1964).
\bibitem{Kadison} R. Kadison, Topology {\bf 3}, supp. 2, 177 (1965).
\bibitem{POVMs} F. Buscemi, G. M. D'Ariano, M. Keyl, P. Perinotti, R.
Werner, J. Math. Phys. {\bf 46}, 082109 (2005).
\bibitem{fbounds} M.A. Nielsen, I.L. Chuang, {\it Quantum Computation and Quantum Information}, Cambridge
University Press (2000).
\bibitem{Uhlmann} A. Uhlmann, Rep. Math. Phys. {\bf 9}, 273 (1976).
\bibitem{positivemaps} E. Stoermer, Acta Math. {\bf 110}, 233 (1963).
\bibitem{unitalqubits} C. King, M.B. Ruskai, IEEE Trans. Info. Theory, {\bf 47} 192
(2001).
\bibitem{unitalqubits2} A. Fujiwara, P. Algoet, Phys. Rev. A {\bf 59}, 3290
(1999).
\bibitem{2convexcp} M.B. Ruskai, S. Szarek, E. Werner, Lin. Alg. Appl. {\bf 347}, 159
(2002).
\bibitem{2convex} V. Gorini, E.C.G. Sudarshan, Comm. Math. Phys. {\bf 46}, 43
(1976).
\bibitem{filtering} F. Verstraete, H. Verschelde, quant-ph/0202124
(2002); F. Verstraete, J. Dehaene, B. De Moor., Phys. Rev. A {\bf
64}, 010101(R) (2001).
\bibitem{2special} K.G.H. Vollbrecht, R.F. Werner, J. Math. Phys. {\bf 41}, 6772
(2000).
\bibitem{Bacon} D. Bacon, A.M. Childs, I.L. Chuang, J. Kempe, D.W. Leung, X.
Zhou, Phys. Rev. A {\bf 64}, 062302 (2001).
\bibitem{quasifree} J. Eisert, M.M. Wolf, quant-ph/0505151 (2005);
 `Gaussian quantum channels', in: {\it Quantum Information with continuous
variables of atoms and light}, N. Cerf, G. Leuchs, and E.S. Polzik
(Eds.) (Imperial College Press, London, 2006).
\bibitem{RG} F. Verstraete, J.I. Cirac, J.I. Latorre, E. Rico, M.M.
Wolf, Phys. Rev. Lett. {\bf 94}, 140601 (2005).
\end{thebibliography}
\end{document}